\documentclass[10pt,preprint]{aastex}
\usepackage{psfig}

\slugcomment{Submitted to the {\it The Astrophysical Journal}}

\newcommand{\lapprox}{{_< \atop{^\sim}}} 

\begin{document}
\def\ale{\mathrel{\hbox{\rlap{\hbox{\lower4pt\hbox{$\sim$}}}\hbox{$<$}}}}
\def\age{\mathrel{\hbox{\rlap{\hbox{\lower4pt\hbox{$\sim$}}}\hbox{$>$}}}}

\shorttitle{GRB~020410 and its Supernova}
\title{GRB~020410: A Gamma-Ray Burst Afterglow
Discovered by its Supernova Light\altaffilmark{1}} 

\author{
Andrew~Levan\altaffilmark{2,3},
Peter~Nugent\altaffilmark{4},
Andrew~Fruchter\altaffilmark{3},
Ingunn~Burud\altaffilmark{3},
David~Branch\altaffilmark{5},
James~Rhoads\altaffilmark{3},
Alberto~Castro-Tirado\altaffilmark{6},
Javier~Gorosabel\altaffilmark{6,3},
Jos\'e Mar\'{\i}a Castro Cer\'on\altaffilmark{3},
Stephen E. Thorsett\altaffilmark{7},
Chryssa Kouveliotou\altaffilmark{8},
Sergey Golenetskii\altaffilmark{9},
Johan Fynbo\altaffilmark{10},
Peter Garnavich\altaffilmark{11}, 
Stephen Holland\altaffilmark{12,13},
Jens Hjorth\altaffilmark{14},
Palle M\o ller\altaffilmark{15},
Elena Pian\altaffilmark{16},
Nial Tanvir\altaffilmark{17}
Mihail Ulanov\altaffilmark{9},
Ralph Wijers\altaffilmark{18},
Stan Woosley\altaffilmark{7}}

\altaffiltext{1}{Based on observations made with the NASA/ESA {\it Hubble
Space Telescope}, obtained at the Space Telescope Science Institute,
which is operated by the Association of Universities for Research in
Astronomy, Inc., under NASA contract NAS 5-26555. These observations
are associated with programs 9074 and 9405.Based partly on
observations obtained at the European Southern Observatory (ESO)
under programme 165.H-0464}
\altaffiltext{2}{Department of Physics and Astronomy, University of
Leicester, University Road, Leicester, LE1 7RH, UK, email: anl@star.le.ac.uk}
\altaffiltext{3}{Space Telescope Science Institute, 3700 San Martin 
Drive, Baltimore, MD 21218, USA}
\altaffiltext{4}{Lawrence Berkeley National Laboratory, 1 Cyclotron Road,
Berkeley, CA 94720, USA }
\altaffiltext{5}{Department of Physics and Astronomy, University of
Oklahoma, Norman, OK 73019, USA}
\altaffiltext{6}{Instituto de Astrof\'{\i}sica de Andaluc\'{\i}a
(IAA-CSIC), Apartado de Correos, 3.004, E--18.080 Granada, Spain}
\altaffiltext{7}{Department of Astronomy and Astrophysics, 
University of California, 1156 High Street, Santa Cruz, CA 95064, USA}
\altaffiltext{8}{NASA/Marshall Space Flight Center, National Space Science
and Technology Center (NSSTC), SD-50, 320 Sparkman Drive, Huntsville, AL
35805, USA}
\altaffiltext{9}{Ioffe Physico-TechniInstitute, St. Petersburg 194021, Russia}
\altaffiltext{10}{Department of Physics and Astronomy, University of
Aarhus, Ny Munkegade, DK-8000 Aarhus C. Denmark}
\altaffiltext{11}{Department of Physics, University of Notre Dame, 225
Nieuwland Science Hall, Notre Dame, IN 46556, USA} 
\altaffiltext{12} {Swift Science Center, Goddard Space Flight Center, 
Code 660,1, Greenbelt, MD 20771-0003, USA}
\altaffiltext{13} {Universities Space Research Association}
\altaffiltext{14} {Niels Bohr Institute, Astronomical Observatory, 
University of Copenhagen, Juliane Maries Vej 30, DK-2100, Copenhagen, 
Denmark}
\altaffiltext{15} { European Southern Observatory, 
Karl-Schwarzschild-Strasse 2, D-85748 Garching bei München, Germany}
\altaffiltext{16} {Osservatorio Astronomico di Trieste, Via
G.B. Tiepolo 11, 34131 Trieste, Italy}
\altaffiltext{17} {Department of Physical Science, University of
Hertfordshire, College Lane, Hatfield, AL10 9AB, UK}
\altaffiltext{18} {Astronomical Institute, University of Amsterdam,
Kruislann 403, 1098 SJ Amsterdam, The Netherlands}

\begin{abstract}
We present the discovery and monitoring of the optical
transient (OT) associated with GRB~020410. The fading OT
was found by {\it Hubble Space Telescope (HST)} observations taken 28
and 65 days
after burst at a position consistent with the X-ray  
afterglow. Subsequent re-examination of early ground based observations
revealed that a faint OT was present 6 hours after burst, confirming the source
association with GRB~020410. A deep non-detection after one week requires
that the OT re-brightened between day 7 and day 28, and further late
time {\it HST}
data taken approximately 100 days after burst imply that it is very red
($F_{\nu} \propto \nu^{-2.7})$. We compare
both the flux and color of the
excess with supernova models and show that the data are best explained by
the presence of a Type Ib/c supernova at a redshift $z \approx 0.5$, which
occured roughly coincident with the day of GRB.

\end{abstract}


\section{Introduction} \label{sec:intro} 

The discovery in 1997 of multi-wavelength counterparts to gamma-ray
bursts (GRBs) \citep{pgg97,cfh97,bre98,fkn97} has led to exceptionally rapid
development in the field. Evidence has accumulated that at least the
GRBs with durations longer than about two seconds \citep{kou93} are
associated with the collapse of massive stars. These events are often
called collapsars \citep{wo93,wo99}.

The apparent association of GRB~980425 with the luminous Type Ic
supernovae 1998bw \citep{98bwiauc} at $z=0.0085$ provided the first
observational evidence of a link between GRBs and supernovae.  However
GRB~980425 was a very unusual event. No apparent optical afterglow was
discovered after the burst; instead, the rising supernova lightcurve
was seen.  The $\gamma$-ray luminosity was substantially lower than
that inferred for many GRBs: the total burst energy was only $\sim
10^{47}$ ergs (if the emission is assumed to be isotropic)
compared with isotropic energy releases for ``classical'' GRBs
of 10$^{51} - 10^{54}$ ergs \citep{bkp01,bkf03}. These significant
differences from other well-studied GRB afterglows raised concern that
GRB~980425 may have been a unique event, making it difficult to reach
firm conclusions about the association of supernovae with other GRBs.

Nevertheless, the SN1998bw/GRB~980425 connection sparked considerable
interest in searching for supernova events associated with other
bursts.  Strong affirmative evidence came from studies of the afterglow of the
nearby GRB~030329, in which the characteristic spectral signatures of
a Type Ic supernova were found at late times \citep{hjorth03,stanek03,
garn03}.
This burst event, at least, appears to have accompanied the
destruction of a massive star---supporting a key prediction of the
collapsar models.

Are all GRBs associated with supernovae?  If so, are the SNe all of
the same type, or is there a diversity of supernova properties? Can we
study these SNe in order to constrain the nature of the GRB
progenitors?  To fully answer this question we will need to find
signatures of supernovae at distances too great to reach with direct
spectroscopy. Supernova spectra peak in the rest-frame optical, so SNe
become hard to observe at high redshift where optical observations
probe rest-frame ultraviolet light. Furthermore, the host galaxy is
often brighter than the supernova.

Most supernovae associated with GRBs have been discovered
only by their contribution to the intermediate time lightcurve of the
GRB afterglow.  Supernovae start faint, and reach a peak days to weeks
later, depending on the supernova type, the progenitor mass, the
envelope structure, the environment density profile and the energy
release.  Conversely GRB afterglows
are bright at very early times, and generally fall monotonically with
little color variation. As the supernova brightens, the fading of the
afterglow can appear to have slowed down, or even been reversed 
with a brightening
of the SNe.  At the same time, the color becomes dominated
by the supernova spectrum, which is typically more red than the power-law
spectral energy distribution seen at early times.  Such ``supernova
bumps'' have now been seen in several GRB afterglows, including
GRBs~980326, 970228, 991208, 011121, and 020405
\citep{bkd+99,cas326,gtv00,cas208,gss03,garn03a,bkp02,pks02,mpp}. 
They are in general adequately
fit by the colors and brightnesses of high-velocity Type Ib/c
supernovae such as SN~1998bw.


Other evidence for a connection between supernovae and GRBs comes from
observations of the galaxies in which GRBs occur. Host galaxies
exhibit very blue colors \citep{ftm99} and strong emission lines (e.g.
[O~{\sc II}],[O~{\sc III}]), indicative of high star formation rates 
\citep{vfk01}, as
expected in models involving the collapse of massive stars, though
perhaps not ruling out models in which GRBs are caused by the rapid
merger of young binary neutron stars. Perhaps stronger evidence comes
from the fact that all but one of the well localized GRBs are associated
with a host
galaxy \citep{bkd+02}, and furthermore the positions of the GRBs 
trace the square of the rest frame UV surface brightness 
(Fruchter et al, in prep). This association suggests a strong 
link between massive stars and GRBs. 

Here we show that the GRB~020410 was likely associated with a
high-velocity supernova of type Ib/c supernova at a redshift $z\sim 0.5$. The source
brightened between ground-based observations taken after one week and
those taken with {\it HST} after one month, our late time observations also
show a red color, typical of SNe. Fitting of the observed
lightcurve implies that the peak absolute magnitude of the underlying SNe
was $M_V \sim -17$, approximately 2 magnitudes fainter than
SN~1998bw. Similar, faint SNe may underlie many long GRBs, but would
remain undetected in typical followup campaigns.

\section{Observations}


GRB~020410 was detected by the {\it BeppoSAX} Wide Field Camera 2 (WFC)
on 10 April 2002 at 10:41:20 UT \citep{gan02a}. 
The Gamma-Ray Burst Monitor (GRBM) 
was switched off at the time of the burst and thus no high energy
($> 28$ keV) data  were recorded by {\it BeppoSAX}. 
However the burst was detected by {\it Konus-Wind} in waiting mode 
(it did not trigger the instrument)\footnote{The detection
of the GRB by {\it Konus-Wind} was first reported by Nicastro
2002 at the ``Gamma-Ray Burts in the Afterglow Era'' conference}. 
In the 19-300 keV band it had
a duration of $\sim 600s$ and was best fit with a powerlaw 
with slope $\Gamma = -1.9$, it had a peak flux of $8 \times 10^{-8}$ 
ergs s$^{-1}$ cm$^{-2}$ and fluence $1.3 \times 10^{-5}$ ergs
cm$^{2}$ in the 19-1270 keV band. The lightcurves in the 19-77, 77-330 and 330-1270 keV 
energy bands are shown in Fig 1. 

{\it BeppoSAX} observations with the Narrow Field Instruments (NFI) 
began 20 hours 
after burst, with a second epoch obtained after 54 hours. These 
observations revealed the presence of slowly fading X-ray point source
\citep{gan02b,gan02c,npg02} with a decay  
best fit as $t^{-0.82}$; \citep{npg04}), the initial error
radius was two arcminutes, however this was further refined to
a 20\arcsec~ radius error box 8 days after the burst. 

Based on both the prompt emission observed by the WFC and the
afterglow properties it was suggested that GRB~020410 was 
an X-ray Flash (XRF)\citep{npg02}. XRFs are defined as burst
which have a higher X-ray fluence than $\gamma$-ray fluence
(Lamb et al, 2003). In order to compare the 
properties of this burst which was observed by {\it Konus-Wind} with
those of the bulk GRB/XRF population (now observed by {\it HETE-2}) 
we calculated the expected fluence in the 7-30 $(S_X)$ and 30-300
($S_{\gamma}$) keV band, 
based on the observed powerlaw slope of $\Gamma = -1.9$. 
We can then calculate the ratio $\log{({S_X}/S_{\gamma})}$. 
X-ray rich GRBs and XRFs are
defined as having $\log{({S_X}/S_{\gamma})}$ $>-0.5$ and $>0$ respectively.
For GRB 020410 we determine $\log{({S_X}/S_{\gamma})} = \log{(5.58 \times 10^{-6} /
1.066 \times 10^{-5})} = -0.28$, which places GRB~020410 in
the X-ray rich GRB category. With a fluence of $1.3 \times 10^{-5}$
ergs cm$^{2}$ (19-1270 keV) the fluence of GRB 020410 lies in the upper quartile
of the bursts reported by Barraud et al (2003). However none of the
bursts with higher fluence are as X-ray rich. Indeed the measured
$\log{({S_X}/S_{\gamma})}$ value of $-0.28$ is actually typical 
of the fainter bursts (for which $\log{({S_X}/S_{\gamma})} =-0.25
\pm 0.07$)  Barraud et al (2003).

A comparison of early optical observations with Digital Sky Survey
images initially failed to find any counterpart \citep{cgc02,kg02}.
However, after the {\it HST} discovery reported here
(and in \cite{flb02}), a reanalysis of
images obtained six hours after the burst at the 0.6m Microlensing
Observations in Astrophysics (MOA)
telescope in New Zealand by Nicastro et al. (2004, in prep) revealed 
a 4-$\sigma$ detection of an OT at
the position of the {\it HST} transient, with a magnitude of R$\sim
21$ (see also Fig 2). This detection confirms the fading source
seen in the {\it HST} images is the afterglow of the GRB, and is
not an unrelated supernova found in the field. 
No radio afterglow was detected to a 4-$\sigma$ limiting flux
of 200 $\mu$Jy at 8.7 GHz \citep{fwb02}. The properties of
the prompt $\gamma$/X-ray emission and of the X-ray afterglow are
discussed in detail in a paper by Nicastro et al. (2004).

The 40\arcsec~diameter error box of GRB~020410 was imaged with
{\it HST}/STIS in 50CCD/CLEAR mode on 8 May and 14 June 2002 as part
of program 9074 (P.I. Fruchter). The dithered datasets were retrieved
from the {\it HST} archive\footnote{http://archive.stsci.edu} after
on-the-fly processing had been applied.  These datasets were then
drizzled \citep{fh02} onto an output grid with pixels half the linear size of
the native pixels (i.e 0\farcs 025 on a side)
 using {\tt pixfrac = 0.7}.  Subtraction of the two
epochs of imaging revealed a transient object approximately 10 \arcsec~
from the center of the refined {\it BeppoSAX} error circle.  The
source was subsequently imaged again with STIS on 22 July 2002 and
18 April 2003. It was also imaged 
with the Advanced Camera for Surveys (ACS) using the Wide Field Camera
(WFC) in the F606W and F814W filters on 24 July 2002 (ACS images as
part of program 9405, P.I. Fruchter). The ACS images were also
drizzled onto an output grid using the same parameters as for the STIS
data. A log of the observations is shown in Table 1.

The {\it BeppoSAX} error circle of GRB~020410 was imaged by the
3.6m telescope at La Silla/ESO seven days after the burst as part of the
GRACE\footnote{Gamma-Ray Afterglow Collaboration at ESO, see
http://zon.wins.uva.nl/\~~grb/grace/} program. Images were obtained in
U,B and R filters, with the R-band exposure being the deepest
observation.  These images were bias subtracted and flat fielded in
the normal way, within {\it IRAF}. Examination of the images reveals 
no evidence for any source at the position of the OT (see below for 
further details). 

\subsection{Photometry and Light Curve}
In order to create a lightcurve free from host galaxy contamination 
we use our late time STIS image of the field of GRB~020410, 
obtained 373 days after the burst as a template
to subtract from the earlier data (see fig 2). We aligned each frame using
8 point sources in common to each image (and the IRAF task {\sc
geomap}) and then drizzled the data onto a common reference frame
The accuracy of this alignment is $\sim$ 0.2 drizzled STIS 
pixels (or 0\farcs005).
We use a small
(0\farcs 1) aperture to determine the flux from the afterglow.
We then perform an aperture correction in order to allow for
the light from the afterglow which is not included in our
measurement aperture. This correction is determined via comparision
with two isolated point sources in the STIS field which exhibit 
a similar color to the afterglow (the colors are determined from
the ACS data). We then perform aperture photometry on each of these
stars using  0\farcs1 and 1\arcsec~ apertures. Having corrected for
the light within the 1 \arcsec aperture we then correct to the
light in an infinite aperture assuming $F_{\infty} = 1.096 \times F_{1''}$.

We determined the zeropoint of the ESO 3.6m image by
comparison to stellar sources in common to both the 3.6m and ACS
images. To do this we measured the observed
color of stellar sources in F814W and F606W and fit a single powerlaw
between the two. We then extracted a Johnson R magnitude from the data
by assuming this spectral slope in {\sc IRAF/synphot}. Using this
method we determine a 3$\sigma$ upper limit of R=24.10 at the position of 
the afterglow. Our {\it HST} images
(see Fig 3) show that the OT is offset from a nearby galaxy by
0\farcs 65. These two sources are blended at the resolution
of our ground based observations, therefore their integrated magnitude
was R$\geq$ 24.10, 7 days after the GRB. We derive a magnitude for the
nearby galaxy based on its measured magnitude in each of
F606W and F814W and its F606W-F814W color of 0.95. This is
converted to an $R$-band magnitude via {\sc IRAF/synphot}, for this
galaxy we determine $R$=24.45. We then
measure the flux on our ground based image within a 1 \arcsec
aperture, which contains both this galaxy and the position of the
optical transient. Their integrated magnitude is fainter than 
$R$= 24.10, using this, and the known magnitude of the nearby galaxy
we derive a  3-$\sigma$ upper limit on the magnitude
of the afterglow as R$\geq$ 25.0 at $t_b$ =6.8 days.
We have also photometrically calibrated the field of 020410 with 
respect to the USNO-A2 catalog. Using 10 sources within 2 \arcmin of
the GRB position we determine a margnially deeper 3-$\sigma$ upper
limit of $R=24.21$, in order that our estimates remain conservative in
what follows we use the marginally shallower limit implied by our {\it 
HST}
observations.

In order to determine a ``Johnson/Cousins'' $V$ or $R$ magnitude from
our broadband STIS/50CCD
image it is necessary to assume a spectrum of the object. The early
time color of the object is poorly constrained since it is only
present in a single $R$-band image. However given the predicted rate
of decline of the OT we believe that there is minimal contamination from the
OT at the time of the first {\it HST} observation and hence we determine
$K$-corrected magnitudes from the assumed spectra of the SN (see
below) and the observed number of counts. The magnitudes are listed in
Table~\ref{tab:photdata}.

The position of the transient is determined from two USNO CCD Astrograph
Catalog (UCAC) stars which lie in the field of the ACS image. The mean
position based on the two available images (F814W and F606W) is RA=
22$^h$ 06$^m$ 32.18$^s$ Dec =-83$^d$ 49$^m$ 27.81$^s$ (J2000) with an accuracy
of $\sim 0\farcs
2$.  This position is offset from the center of the {\it BeppoSAX} NFI error
box by approximately 10\arcsec, fully consistent with the 40\arcsec~
diameter 90\% error box \citep{npg02}.

The magnitude of the transient at the time of our first STIS
observation is $R = 24.95$, marginally brighter than the limit
obtained at day 7, implying that the transient re-brightened in
this period. 
Based on the ESO non-detection the mean decay rate between
the first detection and the non detection at one week
is $ > t^{-1.1}$, somewhat faster than the observed X-ray
decay flux of $t^{-0.82}$ (Nicastro et al. 2004). Were the slopes
identical (which would be the case if they both lay on the
same section of the fireball spectrum (i.e. were not separated by
the cooling break)) then a break would be implied between the
second {\it BeppoSAX} observations and those with ESO. This would not be
unexpected since the majority of GRB afterglows have average decay
rates faster than $t^{-1}$ over the first week, however the lack
of optical coverage over this periods means that strong conclusions
cannot be reached.

The magnitude of the OT in the final ACS epoch is determined using
aperture photometry. A small ($0\farcs 1$ radius) aperture is used to
minimize the effect of the underlying host, which will likely have
a different spectrum for the GRB afterglow and therefore substantially
change the observed color. The magnitude is subsequently corrected
to the true magnitude by comparison to a point source in the image. 
The observed color using this small aperture is F606W-F814W = 1.1 $\pm 
0.05$,
corresponding to a spectral slope of $\nu^{-2.7}$. The use of a 
larger aperture (0\farcs 25) resulted in a color of F606W-F814W = 0.6
$\pm 0.09$. 
This change is due to the contamination of the measurement by host galaxy
and implies that the host is very blue; typical of GRB hosts in
general \citep{ftm99}.

The lightcurve is shown in Figure~\ref{fig:lc}, with the early 
detection in New Zealand,
from Nicastro et al. (2004) and the GRACE data, along with the points
from STIS. We note that the first detection of the OT by STIS is
marginally brighter than the limit implied by the ESO 3.6m after 7
days, indicating that the object re-brightened between the two
observations.

\subsection{Host Galaxy}
Our final STIS image is taken 373 days post burst. At this stage we 
expect minimal contamination from either the GRB afterglow or
associated supernova. Our STIS image shows an extended source
underlying the position of the afterglow, most likely the host galaxy
of GRB~020410. Its association with the galaxy $\sim$ 1\arcsec~to the
southeast in unclear. It has an V(AB) magnitude of 
27.9 $\pm$ 0.2, which
is typical of that expected among GRB host galaxies which have a range of
magnitudes from $21 < R < 30$ \citep{hf99,le03}. It absolute terms
at $z=0.5$ (our preferred redshift, see discussion below), the host
would have an absolute magnitude of
$M_V$ = -14.3, placing it amonst the faintest GRB hosts, two
magnitudes fainter than the SMC, but comparable to some local 
dwarf star forming galaxies such as IC 1613.
Projecting
the position of the GRB onto this image we find that it lies
offset from the centroid of the underlying host galaxy by approximately 
0\farcs 03 $\pm$ 0\farcs01. At $z=0.5$ the GRB would have occured within
200pc of the galaxy centre.
Using a large aperture (0\farcs 25 radius) on our ACS images we 
find that F606W-F814W = 0.60 $\pm$ 0.1. At the epoch of the ACS
observations there
remains contamination from the supernova (which is red,
with (F606W-F814W)$_{SNe}$=1.1 $\pm 0.06$). We can estimate the color of the
host galaxy by subtracting the contribution of the supernova
within the 0\farcs 25 aperture. We determine the contribution
of the supernovae within this aperture by correcting the flux
measured in a 0\farcs 1 aperture to that in a 0\farcs 25 for
a point source, this is done by using a stellar source in our
image which exhibits similar colors to the SNe, we also checked this
value by comparing it with the tabulated ACS aperture corrections
within {\it synphot} and found agreement to better than $0.02$ mags.
The host galaxy magnitude is then $F_{host} = F_{obs} - F_{SN}$.
This results in AB magnitudes of 
28.50 $\pm$ 0.10 in F606W and 27.99 $\pm$ 0.11 in F814W.
Which results in a color of F606W-F814W = 0.51 $\pm 0.15$.
These magnitudes underestimate the true brightness of the galaxy
since the host galaxy immediately under the supernova has been
subtracted.

The relationship between the host galaxy directly underlying the
position of the GRB, and the brighter galaxy, offset 0\farcs 6 to the
south-west (see Fig 3.) is unclear. There may be no physical
association, however if there is then the GRB occured offset from
the nucleus of this galaxy by 3.6 kpc, unusual for GRB events (this 
would lie well outside the half light radius). However in this case
the absolute magnitude of the host system would be $M_V \sim -18$,
similar to the LMC.

\section{Supernova lightcurve fitting}

In order to determine what, if any, supernova lightcurves could fit
the optical photometric data we constructed several spectral templates
of the various supernova sub-types in order to perform the necessary
$K$-corrections to the data. The templates used for the SN~Ia and
SN~II lightcurves can be found in \citet{nugent_kcorr} and
\citet{hdf_gnp99}, respectively. For the SN~Ib/c subtype, we created a
new template based upon the spectra of SN~1998bw \citep{98bw_spect}
made available to us through the SUSPECT database \footnote{see
  http://tor.nhn.ou.edu/\~~suspect/index.html}. These spectra and their
temporal evolution are similar to those of SNe~1997ef and 2002ap, two
other well know high-velocity SN~Ib/c.

In Figure~\ref{sne1a} we plot the lightcurves of a normal SN~Ia over
the redshift range $0.25 < z < 1.25$ as they would appear in the STIS
50CCD clear filter. It is clear from this figure and the data in
Table~\ref{tab:photdata} that the only possible scenario in which the
OT could be associated with a SN~Ia is at a redshift greater than
1.0. SNe~Ia at lower redshifts could only approach the magnitudes seen
for this event at very late times during the tail of the supernova
lightcurve. At these epochs the decay of the lightcurve is nicely
modeled by a decline of 0.017 magnitudes per day in the restframe $B$- or
$V$-band divided by the factor $(1+z)$ over a period of several
hundred days  \citep[for example]{cap_lc}. However, this is not a good match
for the lightcurve of the OT as it declines by 0.05 mag between days
28 and 65 and 0.02 mags between day 65 and 102.

As seen in Figure~\ref{sne1a} a SN~Ia at a redshift of 1.0 can nicely
reproduce the STIS lightcurve if the first {\it HST} data point was
taken near maximum light. However, this possibility can be rejected
with consideration of the ACS photometry in the F606W and F814W filter
taken at day 104. The observed color was $\approx$ 1.1 mag.  However,
for SNe~Ia at this redshift and late-time epoch, the flux in the
restframe ultraviolet has practically vanished and thus the flux in
the F606W filter would as well. Based on our templates the color for a
SN~Ia with $z \ge 1$ at these epochs would be $>2.2$ between these
filters. Thus a SN~Ia at any epoch and redshift is not able to
reproduce all the observed data for the OT.

As for SNe~II the mere shape of the lightcurve rules out plateau-like
events. The only possible fit to a linear-like event, based upon the
colors, would be at very low redshift, $z < 0.2$, as seen in
Figure~\ref{sne2l_a}. However, at these redshifts the intrinsic
brightness of the supernova would be $M_B \approx -14.0$. This is
unheard  of for linear events which are usually representative of the
brighter portion of the SN~II population such as SN~1979C
\citep{branepm,sn79cdev}. Further Richardson et al. (2002)
find that none of their sample of 16 SN II-L events have
$M_B > -16$. For a full discussion of supernova absolute
magnitudes see \citet{rich02}. In addition, as seen in
Figure~\ref{sne2l_b}, the fit to an average linear template at these
redshifts is quite poor.

The only good fit to the OT based on an underlying supernova is seen
in Figure~\ref{97ef_lc}. Here we have shifted the high velocity SNe~Ic
1997ef \citep{iw00}, 1998bw \citep{98bwiauc,mc99} and 2002ap 
\citep{pan03,gy02} to a redshift of $z=0.5$ with the explosion
date set to be coincident with the observation of the GRB signal from
020410 (cosmology of $h=0.7$,$\Omega_M=0.3$ and
$\Omega_{\Lambda}=0.7$). All three supernovae follow a very similar
spectral evolution to the template so that differences in the
$K$-corrections between the various supernovae will be negligible. The
observed lightcurve for the OT is nicely reproduced as is the color at
day 104. Both SNe~1997ef and 2002ap have peak magnitudes of $M_V
\approx -17.4$, which is where the OT is found to lie given our choice
of cosmology. SN~1998bw is more than 2 magnitudes brighter than 
SN 1997ef and SN 2002ap, so we have shifted down the peak of its lightcurve
correspondingly. Some variation in the lightcurve shape and decline on
the tail between these supernovae is seen, demonstrating that within
the expected variability of these types of SNe~Ic we would find a good
match to the lightcurve of GRB~020410

In Figures~\ref{grfit}~and~\ref{light} we demonstrate the constraint on the redshift by
the data. Assuming that the explosion date is coincident with the date
of the GRB detection, we plot the brightness (blue) and color (black)
of a high-velocity SN~Ib/c as a function of redshift along with the
{\it HST} observations. Due to the steepness of the color--redshift
relationship, the only reasonable redshift range is $0.45 \lapprox z
\lapprox 0.6$. The resulting peak brightness, given this redshift range,
is $M_V \approx -17.4$. The constraint on redshift is shown in figures 8
\& 9.

We also compared the data to a more normal set of SNe~Ib/c such as
SNe~1993J (IIb) and 1994I (Ic). These supernovae are characterised by much
lower expansion velocities, indicative of less energetic
exposions. They also have significantly more rapid declines
after maximum light, although both SN 1993J and 1994I reach a
peak brightness comparable to SNe 2002ap and 1997ef. Given color 
information for the OT at the earlier
epochs we would have been able to distinguish between these supernovae
and their high-velocity counterparts quite easily since their
colors before, and at maximum show some variety. However,
given our choice of filters and the corresponding uncertainties in the
observations, the late-time colors for all these supernovae are
indistinguishable. The fact that SNe~1993J and 1994I decay much more
rapidly in the optical \citep{lewis,rich} than the OT could be used as
evidence against more normal SN~Ib/c being associated with this OT.
However, given the diversity in both the rise-times and decline rates
of core-collapse supernovae, we do not think this difference can be
used as strong evidence towards an exclusive association of this OT
with a high-velocity SN~Ib/c. The strength of our redshift
constraint ($0.45 \lapprox z \lapprox 0.6$) is a product of both
color and lightcurve shape. We believe this is resonable as the 
local sample of type Ib/c supernovae (SN 1994I, 1997ef,1998bw and 2002ap) 
have effectively identical late time colors. At present these
are the best models spectral models available, but the sample is still
small at present. The caveat to this result is therefore that should
a larger sample of SN Ib/c spectra reveal differing late time ($>50$
day) colors then the redshift range (and hence peak magnitude) 
that we determine could be significantly affected.

The potential detection of a supernova associated with GRB~020410
was largely possible due to the intrinsic faintness of the
afterglow. For example \citet{pks02} place constraints on the
magnitude of an underlying supernovae to GRB~010921 as 1.34 magnitudes
fainter than SN~1998bw. However the absolute magnitude of SN~1998bw
was $M_V = -19.35$ \citep{98bwiauc}, while both SNe~1997ef and 2002ap were
in excess of 2 magnitudes fainter. Hence, if we assume that the intrinsic
luminosity of the supernova underlying GRB~020410 was similar to
that of the latter then it provides an indication that substantially
less optically luminous supernovae can also give rise to GRBs, and
also that supernovae of this brightness may be present in the
afterglows of low redshift GRBs for which no definitive supernova
signature can be found.

Supernova are not the only possible explanations for the late time
bumps seen in GRB afterglows. \cite{eb00} consider the possibility 
that the bumps may be due to dust echos. In this model the late time 
bump is caused by re-radiation, or reflection of the early optical 
afterglow from dust which lies beyond the sublimation radius of the
burst. However, for a dust echo to
have the same appearance as a supernova, it must be a somewhat
fortuitous event. It has to mimic both the color and the time of
peak emission of a supernova. In all cases where a re-brightening of
the lightcurve has been observed it can be well fit by an underlying
supernova which occurred approximately temporally coincident with the
GRB, making dust echoes unlikely. Furthermore using the excellent
temporal and spectral sampling of the afterglow of GRB~970228,
\cite{r01} demonstrated that dust echoes were not a valid mechanism
for the production of the observed bump.


\section{Comparison with other GRBs}
The implied magnitude of GRB~020410 at 1 day is R$\sim 22.5$.
This places GRB~020410 as one of the fainter GRB afterglows
discovered thus far. Its magnitude at one day is similar to
that of GRBs 980329, 980613, 000630, 020124, 021211 and 030227 
\citep[respectively]
{gor99,fjg01,hj02,bkb02,fps03,cas227}. However its decay rate (at least
in the X-ray) was $\sim t^{-1}$, typical for GRBs in general,
implying that GRB~020410 ``started faint'', rather than faded
fast as seen for GRBs 020124 and 021211. The afterglow could
easily have remained unidentified (and the burst classified as
``dark'') were it not for the {\it HST} observations taken 28 days
after the burst. This is apparently consistent with the conclusions of
Fynbo et al.\ (2001) that many bursts which are classed as optically
dark are in fact simply missed since current searches are neither
sufficiently rapid nor deep enough to discover the OT. In any case, this
burst could certainly be classified among the ``optically dim'' GRBs.

Some models of dark and optically faint GRBs attribute their optical
faintness to extreme extinction in their host galaxies, as may be
expected if they are quite dusty. This would be typical of low
redshift starbursts. To date the exact nature of host extinction and
its effect on the bursts is unclear. For example in the well studied
case of GRB 010222 (\citet{gtb03}) the
multi-band optical data is well fit by a very low extinction, somewhat
in conflict with the SCUBA observation implying the galaxy is a ULIRG
with $A_V \sim 5$ \citep{fbm02}. 
Another clear disagreement comes from the dark GRB~000210, where
a high $N_H$ was measured from the X-ray afterglow \citep{piro02}
and sub-mm observations imply a high dust content \citep{bar03,ber03}
whereas optically its host galaxy shows $A_V \sim 0$ \citep{gor03a}. Similarly dust 
extinction has been suggested as an explanation for the very red
afterglow of GRB~000418 \citep{ksm00} and the non-detection of the
afterglow of GRB~020819. However the host galaxy extinction in the
case of GRB~000418 is, similarly to GRB~000210 very low ($A_V \sim
0.2$; \cite{gor03b}). Only in the case of GRB~030115 (Levan et al, in
prep; Lamb et al, 2003) does a very red afterglow appear to be 
located in a very red (and highly extincted) galaxy. 
If the optical faintness of the GRB 020410
afterglow were due to extinction in the host, this may explain its
observed red color: the presence of a dusty host would significantly
redden the normally blue ($\sim \nu^{-1}$) afterglow spectrum, whilst
(as with the cases above) the global (host galaxy) colors could 
remain very blue. However, the re-brightening of the afterglow 
of GRB~020410 cannot be explained simply by a dusty host galaxy (unless the
dust echo model is also inferred), further the observed colors of the
SN are not consistent with those of SN 1998bw with significant
extinction (with 2 magnitudes of additional extinction SN~1998bw
would have F606W-F814W = 2.0) so extinction is not a valid
method of obtaining both a faint afterglow and faint SNe. We
therefore believe there is unlikely to be significant extinction
along the line of sight to the GRB.

Zeh, Klose \& Hartmann (2004) have carried out a survey of all low-z GRBs
and find that
their intermediate time afterglows can be fit with SN~1998bw-like
templates with a modest dispersion in
their peak luminosities and rise times. The peak luminosity of a type
Ic supernova is determined by its Nickel
yield. A fainter SNe corresponds to a lower Nickel production. 
The faintest SNe within the Zeh, Klose \& Hartmann (2004) 
sample (for which there is a positive detection) is $\sim$ 40 \% as bright
as SN~1998bw. The supernova associated with GRB~020410 is only $\sim$ 15
\% as bright and is as such the faintest SNe yet seen associated with a
GRB, this would correspond to a Nickel yield of $\sim 0.05$
M$_{\odot}$. There are also now three cases of burst whose "bump" spectra and light
curves may be better fit with SN~1994I-like supernovae (GRB~021211, Della
Valle et al,  2003; XRF 030723 Fynbo et al, 2004; GRB~020305, Gorosabel et al,
2004, in prep). However due to the lack of early color information or
spectroscopy we cannot determine if GRB~020410 lies in this class.

\section{Conclusions}
The re-brightening and color of the afterglow of GRB~020410 present
strong evidence for an associated supernova. We have modelled both
the excess and color with various supernova templates and find
that the only acceptable fit is that of a SN~Ib/c at $z \approx 0.5$.

This GRB was unusual in that its initial discovery was done via
photometry taken one and two months after the burst with {\it HST},
after early ground based observations had failed to detect a
transient. Were it not for these deep, late-time observations, this
burst would have been classified as a dark GRB, for which only an
X-ray afterglow was found. These early optical searches were conducted
using small telescopes to depths no greater than the Digitized Sky
Survey. Hence a change in the search strategies to include much
deeper, later-time imaging may be successful in locating OTs
associated with GRBs. However, in order for this method to be
successful, the error boxes will need to be well defined in order to
limit both the time spent searching and to eliminate unrelated,
serendipitously discovered supernovae.

Finally we note that the presence of a SNe component, whose magnitude
is significantly fainter that SN 1998bw (by 2 magnitudes) has dramatic
implications for the searches for similar such signatures in other
afterglows. It is possible that similar supernova could be present in
the afterglows of many long GRBs. For example, the well constrainted
upper limit for an underlying SNe to GRB~010921 of 1.34 magnitudes
fainter than SN~1998bw remains consistent with a SNe of similar
magnitude to that underlying GRB~020410. This stresses the need for
accurate photometry of the late time afterglows of many GRBs if we are
indeed to understand the breadth of the luminosity function of
underlying SNe.

\section*{Acknowledgments}
Support for proposals 9074 and 9405 was provided by NASA through a
grant from the Space Telescope Science Institute, which is operated by
the Association of Universities for Research in Astronomy, Inc., under
NASA contract NAS 5-26555. AJL acknowledges support from the Space
Telescope Science Institute Summer Student Programme and from PPARC,
UK. PEN acknowledges support from a NASA LTSA and ATP grants. This research
used resources of the National Energy Research Scientific Computing
Center, which is supported by the Office of Science of the U.S. Department
of Energy under Contract No. DE-AC03-76SF00098. STH acknowledges
support from the NASA LTSA grant NAG5--9364.This work was
conducted in part via collaboration  within the the Research and 
Training Network ``Gamma-Ray Bursts: An Enigma and a Tool'', funded by 
the European Union under contract number HPRN-CT-2002-00294

\clearpage


\begin{deluxetable}{lcrccrr}
\footnotesize
\tablecolumns{6}
\tablewidth{0pt}
\tablecaption{Observations of the afterglow of GRB~020410}
\tablehead{\colhead{$t_b$ (days)} & \colhead{Inst./Filter} &
\colhead{Exp. Time} &
\colhead{V-mag} & \colhead{V-mag - host} & \colhead{$K_{BS}$} & 
\colhead{$K_{VS}$}}
\startdata
6.895        &ESO 3.6m R    &1800 &$R>25$  & -   & - & -\\
28.0635      &STIS/50CCD    &8392 &25.35  &25.39$\pm$ 0.03 & -0.177& 0.289\\
65.0635      &STIS/50CCD    &8315 &26.91  &27.22$\pm$ 0.13 & -0.308& 0.582\\
102.820      &STIS/50CCD    &8315 &27.45  &28.29$\pm$ 0.41 & -0.285 &0.484\\
372.215      &STIS/50CCD    &8233 &28.69  &-               & - & -\\ 
104.035      &ACS/WFC F606W &3680 &27.81(AB) &-  & - & -\\
104.371      &ACS/WFC F814W &3680 &26.75(AB) &-  & - & -\\
\enddata
\tablecomments{Table of observations of GRB~020410, showing 
the time of each observation, the observed magnitude (or limiting
magnitude). Also shown in the case of the STIS images are the measured
magnitudes of the OT before and after the host has been subtracted and
the relevant K-corrections to correct from Bessel B \& V filters to
that of STIS. These are based on the observed spectrum of SN~1998bw,
which is spectroscopically similar to that of both SNe 1997ef and
2002ap and indeed the recent SN~2003dh (GRB~030329). The
uncertainty quoted is based simply on the uncertainty in the
measured number of counts in the aperture, and does not
reflect the potential uncertainty in the underlying spectrum. }
\label{tab:photdata}
\end{deluxetable}

\begin{figure}
\centerline{
\psfig{file=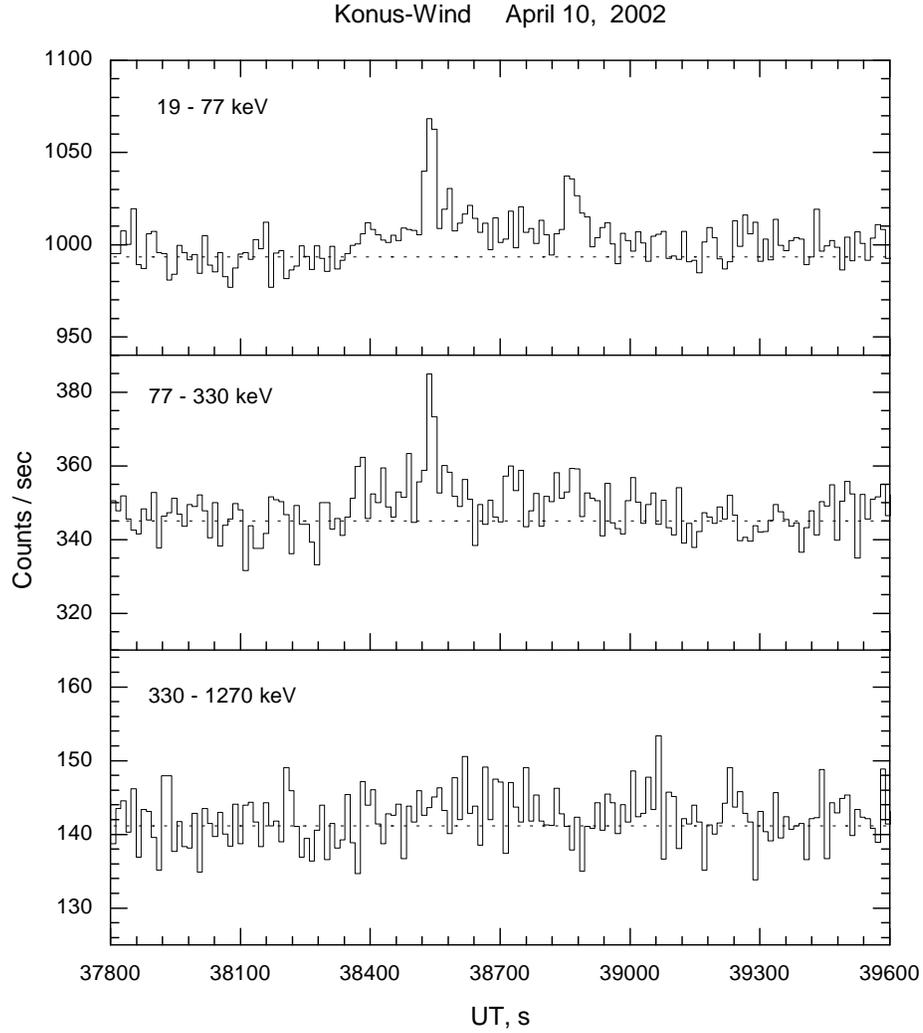,width=6.0in,angle=0}}
\caption{The lightcurve of prompt $\gamma$-ray emssion from
GRB~020410, as seen by {\it Konus-Wind}. The three panels show the
lightcurve in the 19-77, 77-330 and 330-1270 keV. The burst
was only detected in the 19-77 and 77-330 keV range, with
peak flux $8 \times 10^{-8}$ ergs s$^{-1}$ cm$^{-2}$ and
fluxence $1.3 \times 10^{-5}$ ergs cm$^{2}$. The spectrum
is best fitted by a powerlaw of index $\Gamma=-1.9$. }
\label{fig:lc}
\end{figure}

\begin{figure}
\centerline{
\psfig{file=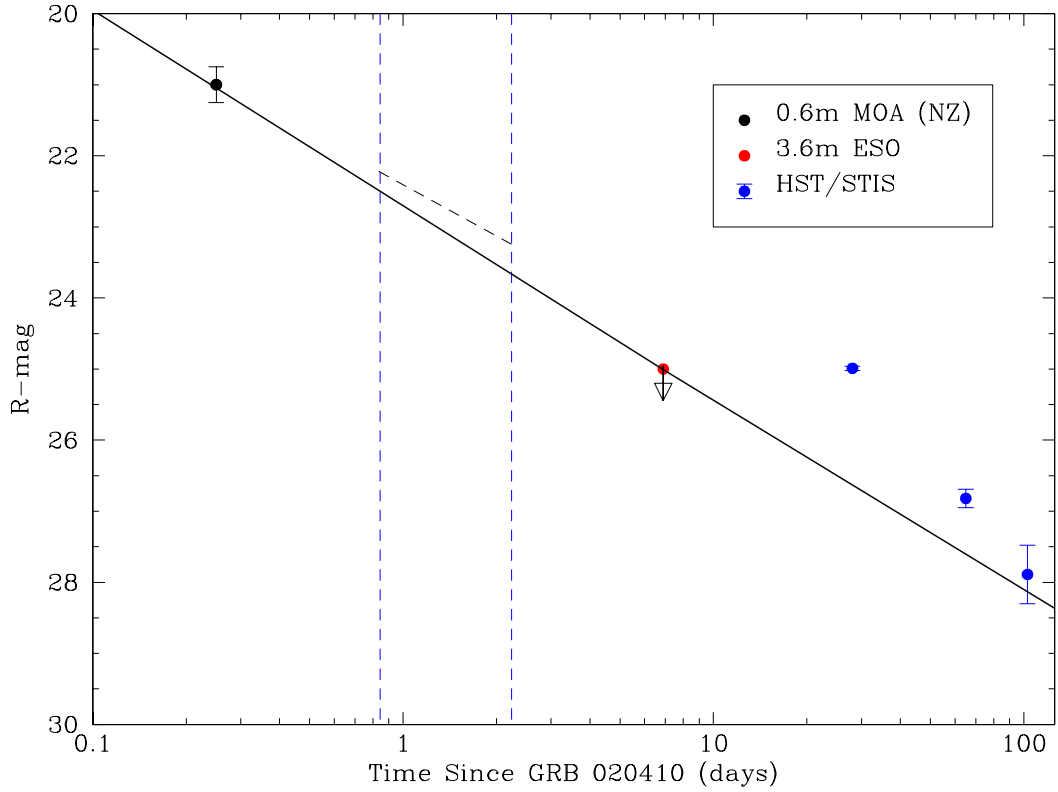,width=6.0in,angle=270}}
\caption{The lightcurve of GRB 020410: The early detection from
the Microlensing in Astrophysics 0.6m telescope (MOA) 
(Nicastro et al. 2004) is shown along with the
non-detection at $t = 7$ days. The solid black line shows the
maximal afterglow contribution under the assumption that it decays
as a single power-law. The true afterglow contribution may lie well
below this line, especially if the jet break occurs between t=1 and
100 days. Also shown are the times of the two {\it BeppoSAX} observations,
and the X-ray slope determined from them. The position of the
X-ray afterglow is arbitary and is for illustrative purposes.}
\label{fig:lc}
\end{figure}

\begin{figure*} 
\centerline{
\psfig{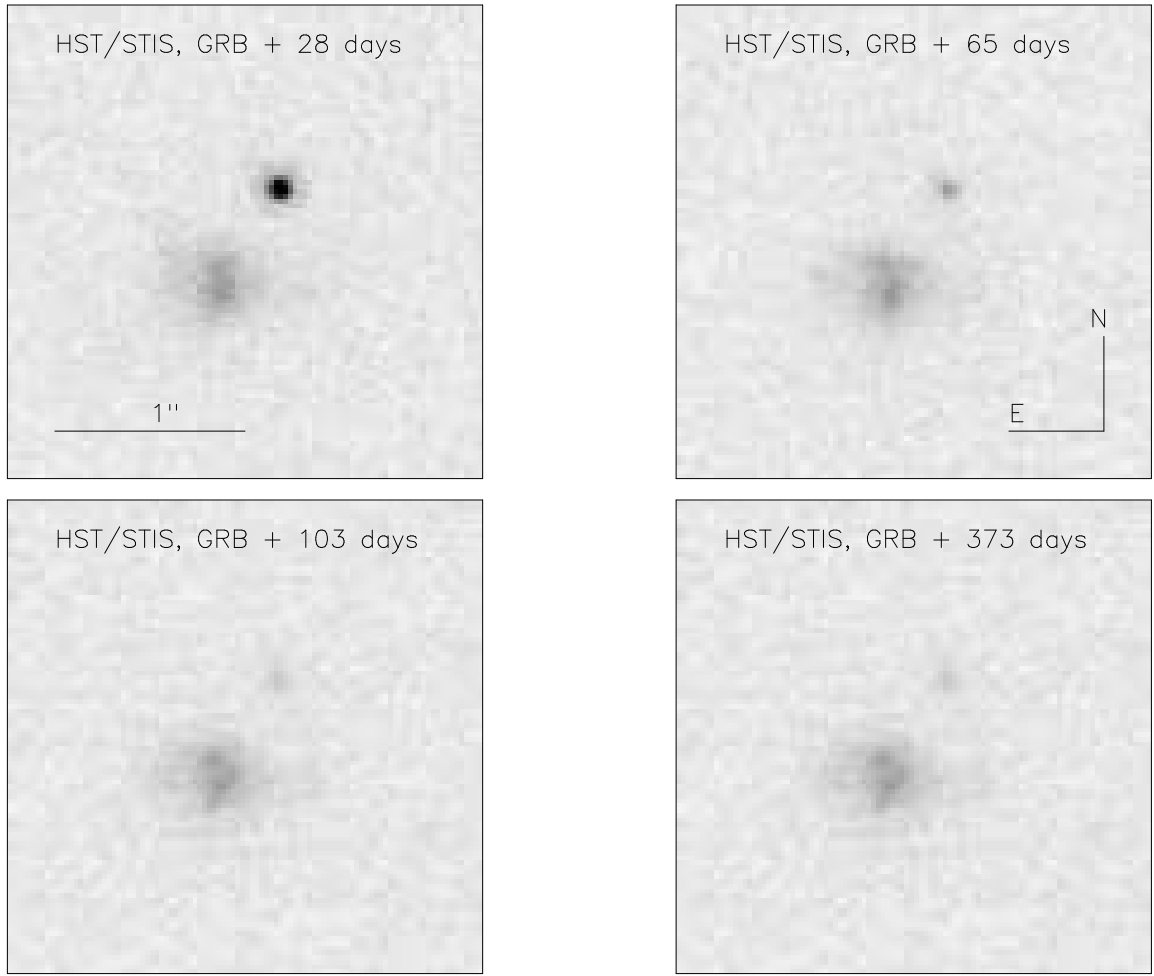}}
\caption[]{Cut down STIS images of the OT of GRB~020410 showing
the fading behavior of the transient over the year between the first
and last observation. In the final image (bottom right) the afterglow
/SNe contributes no light. We interpret the object underlying 
the position of the afterglow as the host galaxy of GRB~020410. This
image is used as a template to subtract from the previous epochs in
order to obtain a lightcurve free from host galaxy contamination.}
\label{fig:stis}
\end{figure*}  

\begin{figure*} 
\centerline{
\psfig{file=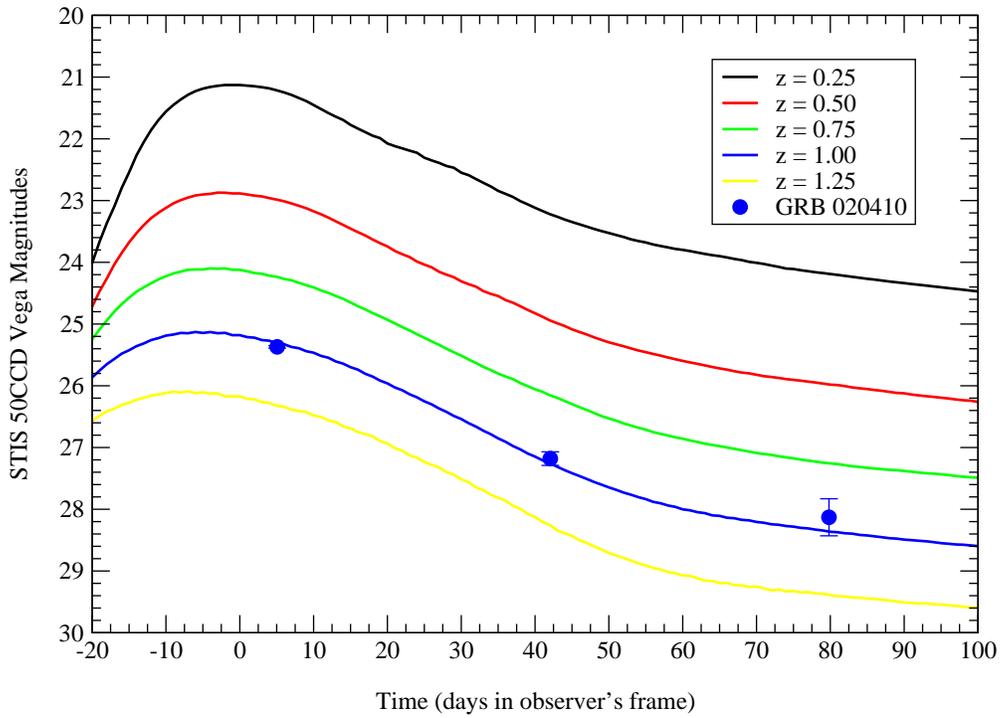,width=6.0in,angle=270}}
\caption[]{The lightcurves of normal SNe~Ia as a function
of redshift. Templates are from Nugent, Kim \& Perlmutter (2002). The
only fit for which a SN~Ia provides a reasonable fit to the data for 
one at a $z=1$. However, the late time
color of the object rules out this possibility.}
\label{sne1a}
\end{figure*}

\begin{figure*} 
\centerline{
\psfig{file=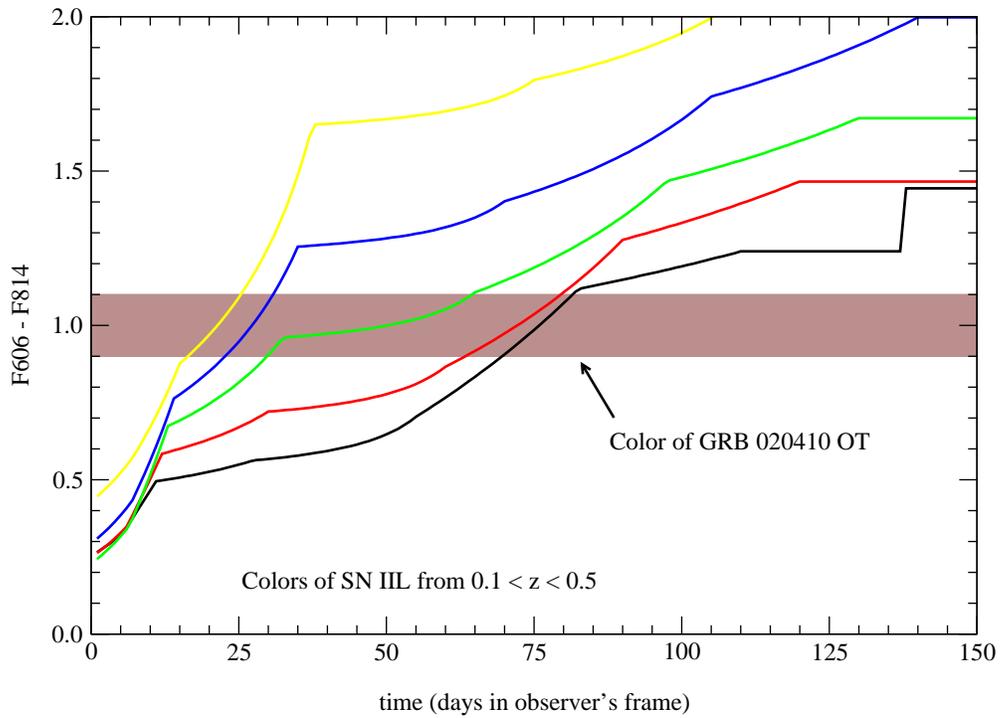,width=6.0in,angle=270}}
\caption[]{The color evolution of  SNe~IIL over a range of redshifts
($0.1$ (black) $\leq z \leq 0.5$ (yellow)). This clearly shows that the observed late
time color of the OT of GRB~020410 can only be explained by a low
redshift ($z \leq 0.2$) event. However, in this case the luminosity
would not fit the observations.}
\label{sne2l_a}
\end{figure*}

\begin{figure*} 
\centerline{
\psfig{file=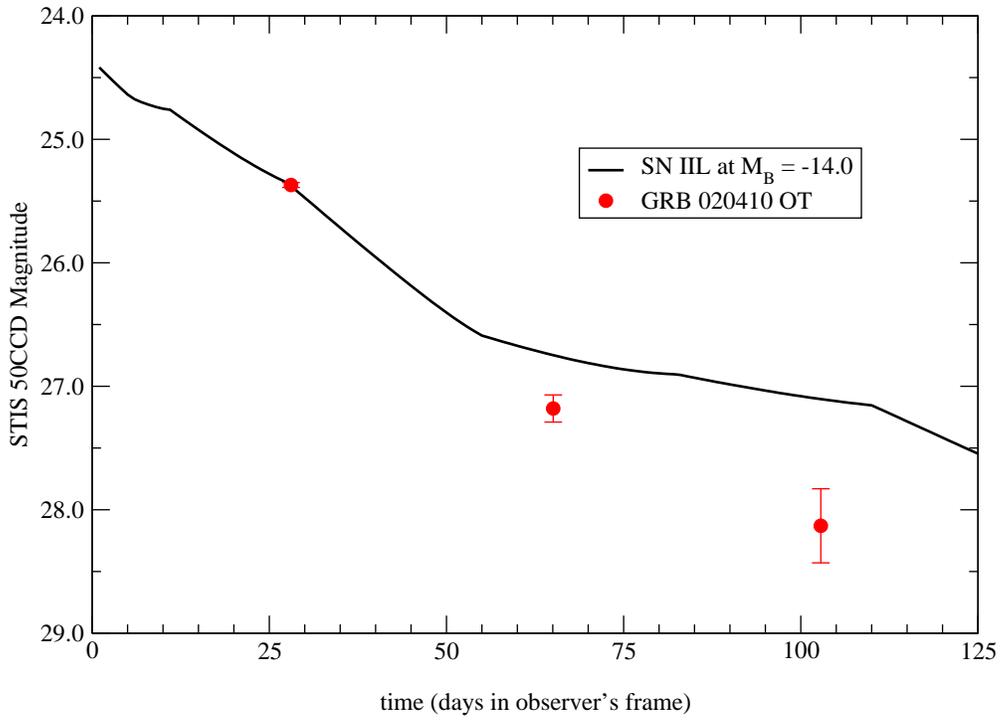,width=6.0in,angle=270}}
\caption[]{A $z=0.2$ SN~IIL, overlayed with the {\it HST} magnitudes. As
can be seen, the late time behavior is poorly modeled by this type of
supernova.}
\label{sne2l_b}
\end{figure*}

\begin{figure*} 
\centerline{
\psfig{file=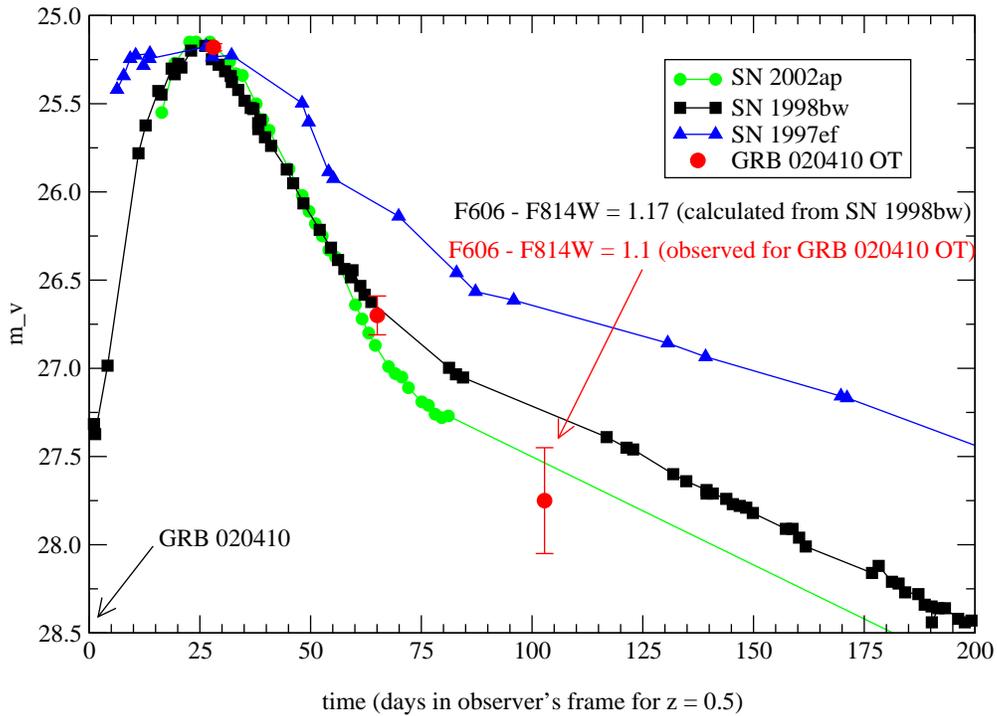,width=6.0in,angle=270}}
\caption[]{A fit to the STIS data points for OT of GRB~020410. The data
are modeled with the lightcurves of SNe~1997ef (Iwamoto et al, 2000), 
2002ap (Gal-Yam, Ofek \& Shemmer, (2002); Pandey et al. (2003))  
and 1998bw (Galama et al. 1998; McKenzie \& Schaefer, 1999),
redshifted to z=0.5. Given our choice of cosmologies, the afterglow
curve is well reproduced by the first two supernovae while SN~1998bw
(which is intrinsically 2 magnitudes brighter) has been shifted down
in brightness in order to examine the diversity in lightcurve shapes
for high-velocity SNe~Ib/c.}
\label{97ef_lc}
\end{figure*}

\begin{figure*} 
\centerline{
\psfig{file=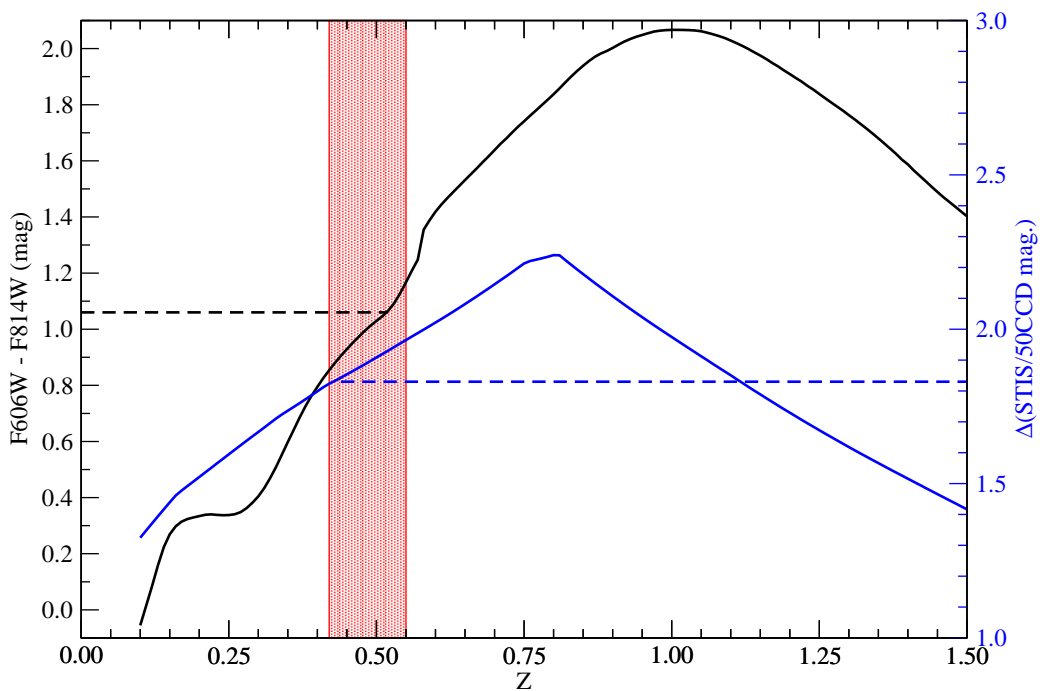,width=6.0in,angle=270}}
\caption[]{Assuming that the explosion date is coincident with the
date of the GRB detection, we plot the brightness (blue) and color
(black) of a high-velocity SN~Ib/c as a function of redshift along
with the {\it HST} observations. 
The $\Delta$(STIS/50CCD) mag value is based on 
the observed $\Delta$(STIS/50CCD) for SN~1998bw and SN~2002ap (which
are very similar over the time frame between the
first two {\it HST} observations, see fig 7). Due to the steepness of the
color--redshift relationship, the only reasonable redshift range is
$0.45 \lapprox z \lapprox 0.60$. The resulting peak brightness is $M_V
\approx -17.4$, assuming the aforementioned cosmology, which is more
consistent with the peak brightness of SNe~2002ap and 1997ef rather
than SN~1998bw. Our color-redshift relation is based on the late times
colors of the SNe ($\sim$ 100 days) where all SN Ic show very
similar colors.}
\label{grfit}
\end{figure*}

\begin{figure*} 
\centerline{
\psfig{file=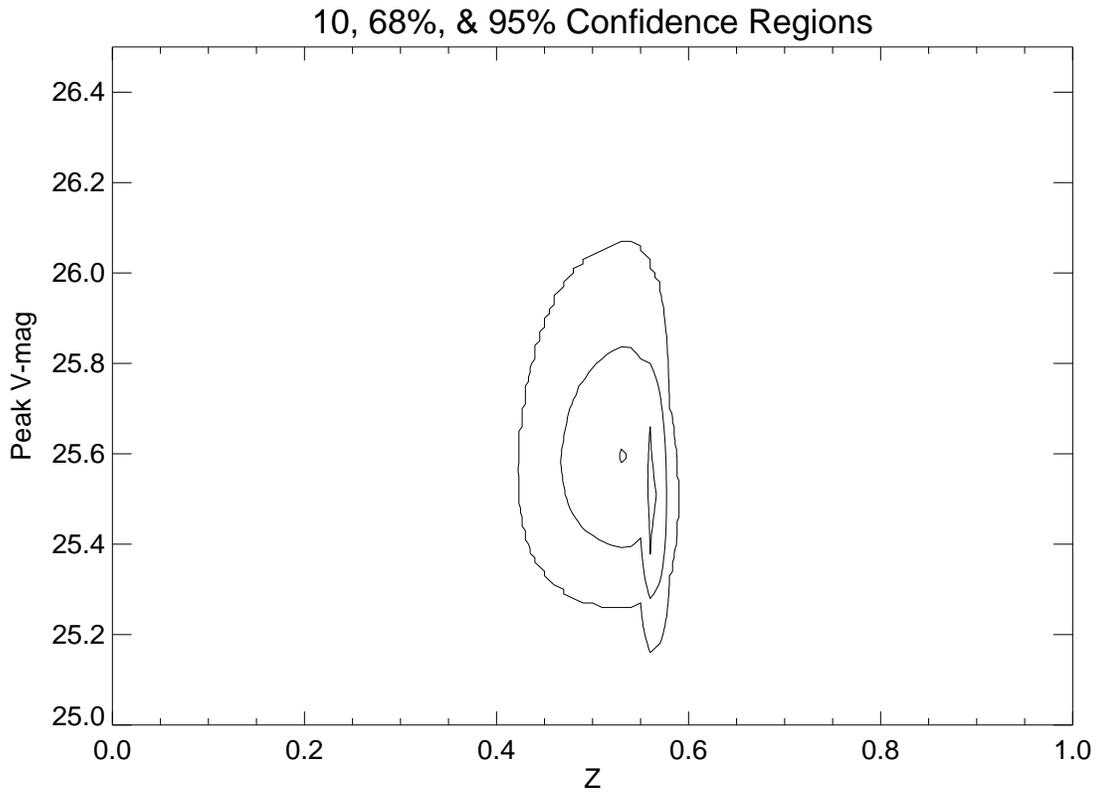,width=6.0in,angle=0.0}}
\caption[]{The redshift confidence range for the supernova
associated with GRB 020410, assuming that the GRB and
SNe are temporally coincident and that the SNe has a
SN 1998bw-like spectrum.}
\label{light}
\end{figure*}

\end{document}